\documentclass[longauth]{aa}

\usepackage{latexsym}		
\usepackage{graphicx}		
\usepackage{rotating}		
\usepackage{natbib}  
\usepackage{savesym}
\usepackage{pdflscape}
\usepackage{amssymb}
\savesymbol{doublespace}
\usepackage{wrapfig}
\usepackage{xspace}
\usepackage{color}
\usepackage{multicol}
\usepackage{mdframed}
\usepackage[draft]{hyperref} 
\usepackage{subfigure}
\usepackage{lscape}
\usepackage{grffile}
\usepackage{standalone}
\standalonetrue
\usepackage{import}
\usepackage[utf8]{inputenc}
\usepackage{longtable}
\usepackage{booktabs}
\usepackage[yyyymmdd,hhmmss]{datetime}
\usepackage{fancyhdr}

\newcommand{\msun}{\ensuremath{M_{\odot}}\xspace}			

\newcommand{\hh}{\ensuremath{\textrm{H}_{2}}\xspace}			

\newcommand{\formaldehyde}{\ensuremath{\textrm{H}_2\textrm{CO}}\xspace}

\newcommand{\methanol}{\ensuremath{\textrm{CH}_3\textrm{OH}}\xspace}
\newcommand{\fivesix}{\ensuremath{5_1 - 6_0$~A$^+}\xspace}
\newcommand{\sevensix}{\ensuremath{7_0 - 6_1$~A$^+}\xspace}

\newcommand{\oneone}{\ensuremath{1_{1,0}-1_{1,1}}\xspace}

\newcommand{\water}{H$_{2}$O\xspace}		

\newcommand{\kms}{\textrm{km~s}\ensuremath{^{-1}}\xspace}	

\newcommand{\um}{\ensuremath{\mu \textrm{m}}\xspace}    

\newcommand{\ammonia}{NH\ensuremath{_3}\xspace}

\def\eqref#1{Equation \ref{#1}}

\renewcommand\arcsec{\mbox{$^{\prime\prime}$}\xspace}

\def\Figure#1#2#3#4#5{
\begin{figure*}[htp]
\includegraphics[scale=#4,width=#5]{#1}
\caption{#2}
\label{#3}
\end{figure*}
}

\def\FigureTwoAA#1#2#3#4#5#6{
\begin{figure*}[htp]
\subfigure[]{ \includegraphics[scale=#5,width=#6]{#1} }
\subfigure[]{ \includegraphics[scale=#5,width=#6]{#2} }
\caption{#3}
\label{#4}
\end{figure*}
}

\newenvironment{rotatepage}%
{}{}

\begin{document}

   \title{High-mass star-forming cloud G0.38+0.04 in the Galactic center dust
   ridge contains \formaldehyde and SiO masers}

   \titlerunning{Cloud C Masers}
   \authorrunning{Ginsburg et al}



\author{
        Adam Ginsburg{\inst{\ref{eso}}},
        Andrew Walsh{\inst{\ref{curtin}}},
        Christian Henkel{\inst{\ref{mpifr}},\inst{\ref{saudi}}},
        Paul A. Jones{\inst{\ref{unsw}}},
        Maria Cunningham{\inst{\ref{unsw}}},
        Jens Kauffmann{\inst{\ref{mpifr}}},
        Thushara Pillai{\inst{\ref{mpifr}}},
        Elisabeth A.C. Mills{\inst{\ref{nrao}}},
        Juergen Ott{\inst{\ref{nrao}}},
        J.M. Diederik Kruijssen{\inst{\ref{mpia}}},
        Karl M. Menten{\inst{\ref{mpifr}}},
        Cara Battersby{\inst{\ref{cfa}}},
        Jill Rathborne{\inst{\ref{csiro}}},
        Yanett Contreras{\inst{\ref{leiden}}},
        Steven Longmore{\inst{\ref{ljmu}}},
        Daniel Walker{\inst{\ref{ljmu}}},
        Joanne Dawson{\inst{\ref{csiro}},\inst{\ref{macquarie}}},
        John A.P. Lopez{\inst{\ref{unsw}}}
        }

\institute{
    {\it{European Southern Observatory, Karl-Schwarzschild-Strasse 2, D-85748 Garching bei München, Germany\\
    \email{Adam.Ginsburg@eso.org}}}\label{eso}
    \and
    {\it{International Centre for Radio Astronomy Research, Curtin University, GPO
    Box U1987, Perth WA 6845, Australia}}\label{curtin}
    \and
    {\it{Max--Planck--Institut für Radioastronomie, Auf dem Hügel 69,
    D--53121 Bonn, Germany}}\label{mpifr}
    \and
    {\it{Astron. Dept., King Abdulaziz University, P.O. Box 80203,
    Jeddah 21589, Saudi Arabia}}\label{saudi}
    \and
    {\it{School of Physics, University of New South Wales, Sydney NSW 2052, Australia}}
    \label{unsw}
    \and
    {\it{CSIRO Astronomy and Space Science, P.O. Box 76, Epping, NSW 1710,
    Australia}}\label{csiro}
    \and
    {\it{National Radio Astronomy Observatory, Socorro}}\label{nrao}
    \and
    {\it{Astronomisches Rechen-Institut, Zentrum für Astronomie der
    Universität Heidelberg, Mönchhofstra\ss e 12-14, 69120 Heidelberg,
    Germany}}\label{mpia}
    \and
    {\it{Harvard-Smithsonian Center for Astrophysics, 60 Garden
    Street, Cambridge, MA 02138, USA}}\label{cfa}
    \and
    {\it{Leiden Observatory, Leiden University, PO Box 9513, NL-2300 RA Leiden,
    the Netherlands}}\label{leiden}
    \and
    {\it{Astrophysics Research Institute, Liverpool John Moores
    University, IC2, Liverpool Science Park, 146 Brownlow Hill, Liverpool L3
    5RF, United Kingdom}}\label{ljmu}
    \and
    {\it{Department of Physics and Astronomy and MQ Research Centre in Astronomy,
    Astrophysics and Astrophotonics, Macquarie University, NSW 2109, Australia}}\label{macquarie}
    }


  \abstract {We have discovered a new \formaldehyde (formaldehyde) \oneone
  4.82966 GHz maser in Galactic center Cloud C, G0.38+0.04.  At the time of
  acceptance, this is the eighth region to contain an \formaldehyde maser
  detected in the Galaxy.  Cloud C is one of only two sites of confirmed high-mass
  star formation along the Galactic center ridge, affirming that \formaldehyde
  masers are exclusively associated with high-mass star formation.  This
  discovery led us to search for other masers, among which we found new SiO
  vibrationally excited masers, making this the fourth star-forming region in
  the Galaxy to exhibit SiO maser emission.  Cloud C is also a known source of
  \methanol Class-II and OH maser emission.  There are now two known regions
  that contain both SiO and \formaldehyde masers in the CMZ, compared to two
  SiO and six \formaldehyde in the Galactic disk, while there is a relative
  dearth of \water and \methanol Class-II masers in the CMZ.  SiO and
  \formaldehyde masers may
  be preferentially excited in the CMZ, perhaps because of higher gas-phase
  abundances from grain destruction and heating, or alternatively \water and
  \methanol maser formation may be suppressed in the CMZ.  In any case, Cloud C
  is a new testing ground for understanding maser excitation conditions.  }


   \maketitle

\section{Introduction}
Masers are important tracers of star formation, shocked gas, evolved stars, and
in other galaxies, circumnuclear disks.  While many masers are common in
the Galaxy and readily detected in other galaxies (e.g., OH, \methanol, and
\water), \formaldehyde has only been detected as a maser in seven locations within
our Galaxy, and so far no instances have been confirmed in other galaxies
\citep{Araya2007c,Mangum2008a}\footnote{\citet{Baan1986a} claimed a maser
detection in Arp 220, but \citet{Mangum2008a} reported that this emission
can be explained by thermal processes.  However, Baan (private communication)
reports that high-resolution observations reveal the emission to be nonthermal.
The debate seems unresolved at present.}.

Most of the \formaldehyde masers detected so far have been observed as part of
dedicated surveys targeting high-mass young stellar objects 
\citep[YSOs;][]{Araya2004a,Araya2007b,Araya2008a}.  Despite concerted effort, very few
new masers outside of Sgr B2 \citep{Whiteoak1983a,Mehringer1994b} have been found
since their initial discovery by \citet{Forster1980a}. All of the known
\formaldehyde masers are associated with regions of high-mass star formation
\citep{Pratap1994a,Araya2004a,Araya2007b,Araya2008a}.

The pumping mechanism of the \formaldehyde \oneone maser is not yet understood.
A radio continuum pumping mechanism was proposed by \citet{Boland1981a} and
later \citet{van-der-Walt2014a}, but the lack of bright radio continuum sources
near some of the detected \formaldehyde maser sources means that this mechanism
cannot explain all of the observed masers \citep{Mehringer1994b,Araya2008a}.
\citet{van-der-Walt2014a} ruled out infrared pumping, but suggest that
collisional pumping may be a viable mechanism.  In the
\citet{van-der-Walt2014a} framework, high amplifications $>20$ are not
possible, so additional physical mechanisms must be in play for the brightest
\formaldehyde masers.

SiO masers are common toward oxygen-rich evolved stars, namely long period
variables (Mira stars) and red supergiants \citep[see,
e.g.,][]{Deguchi2004a,Verheyen2012a}, but extremely rare toward star-forming
regions, with only three known \citep{Zapata2009c}.
In the few regions where they have been detected - W51 North, Sgr B2 (M), and Orion KL - 
they closely trace the location of the high-mass YSO, likely pinpointing
the base of a high-velocity outflow \citep{Goddi2015a}.

Cloud C, G0.38+0.04, is one of the high-column-density clouds along the central
molecular zone (CMZ) dust ridge \citep{Lis1999a,Immer2012a}.  It is notable for
containing the brightest 70\,\um point source along that ridge and the third
brightest (after Sgr B2 and Sgr C) along the \citet{Kruijssen2015a} orbit
\citep{Molinari2011a}.  It is not detected at 8 \um with Spitzer
\citep{Yusef-Zadeh2009a} and is therefore  unlikely to be an evolved star, but
it is associated with extended 4.5\,\um emission that is generally observed to
be associated with molecular (\hh) outflows \citep{Chambers2011a}. It is among
the most centrally condensed millimeter sources in the CMZ.  With a mass in
the range 150-2000 \msun, depending on the assumed temperature, it may contain
a single proto-O-star or a proto-cluster (from the SMA-CMZ survey; Battersby,
Keto, et al in prep, Walker et al. in prep).

In the following Letter, we present the serendipitous detection of a
\formaldehyde maser and corresponding new detections of SiO masers in
G0.38+0.04.

\section{Observations}
ATCA observations were performed in 2015 as part of a larger survey of the
CMZ. Observations were conducted on
April 2 and 13, May 11, August 12 and 13, and September 1 and 4  in the
H214, 6A, 1.5C, H75, H75, EW352, and 750B arrays, respectively. The same
spectral setup was used for each array configuration, which included
observations of 14 spectral lines between approximately 4 and 8 GHz. One of our
main target lines is the \oneone transition of \formaldehyde at 4.82966 GHz.
The zoom window at the \formaldehyde frequency yields a channel resolution of
1.9\,\kms over a velocity range 3969\,\kms. The sensitivity of the observations was
$\sigma=2$ mJy/beam in each 1.9\,\kms channel. We assume in this paper that the
absolute positional
uncertainty of the observations is typically 0.4\arcsec but no worse than
1.0\arcsec, based on previous ATCA observations \citep{Caswell2009a}.

\section{Analysis}
\label{sec:othermasers}
We detect spatially and spectrally unresolved \formaldehyde \oneone emission in
one narrow line ($\sigma<1.3$ \kms, below the instrument resolution) at
$v=36.7$ \kms with an amplitude of 235 mJy/beam, where the restoring beam is
$4.84\arcsec\times1.49\arcsec$.  This translates to a brightness temperature of
1700 K.  Molecular emission lines with this brightness are generally, not
observed in thermally excited regions, so it indicates that there is maser emission.

\FigureTwoAA
{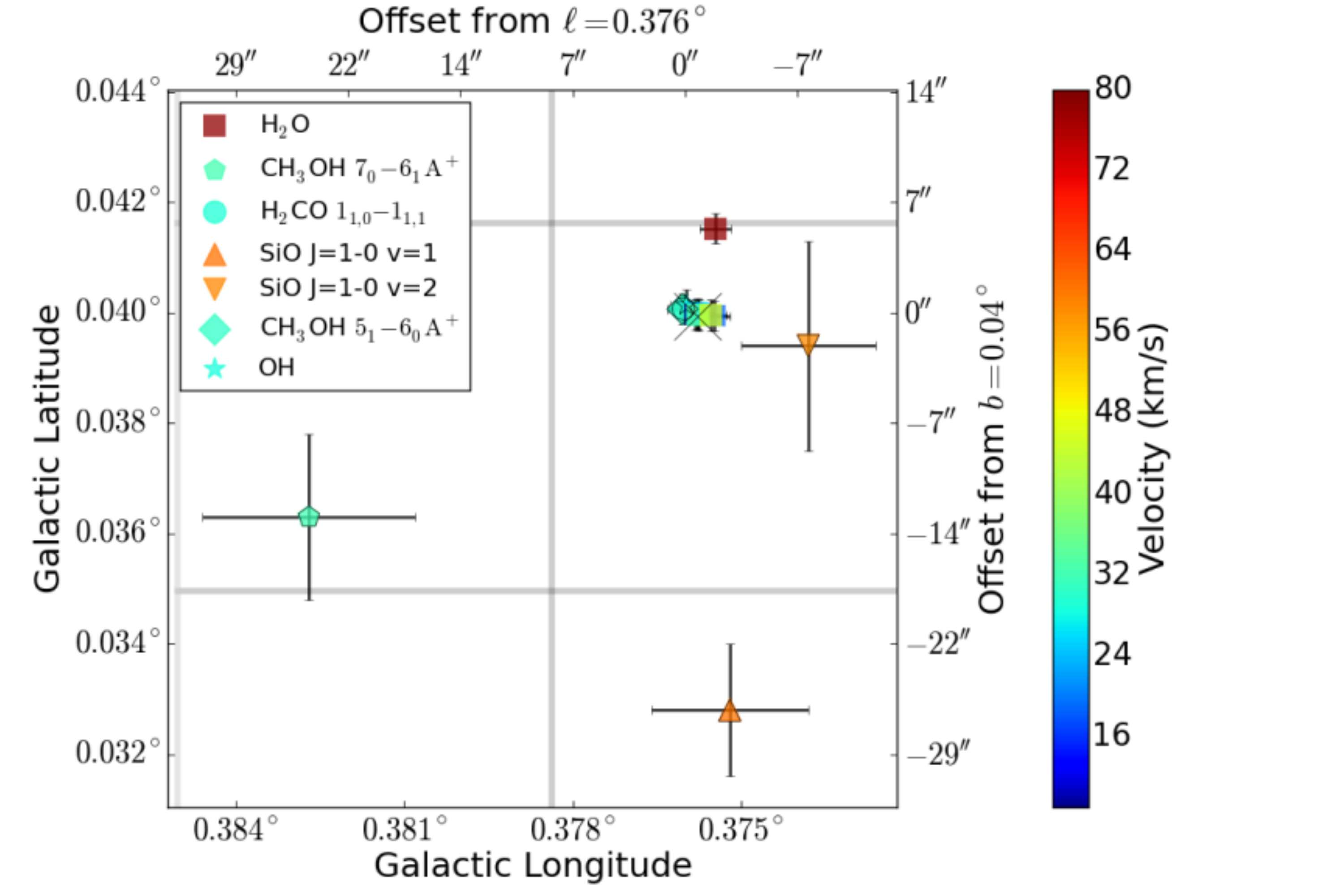}
{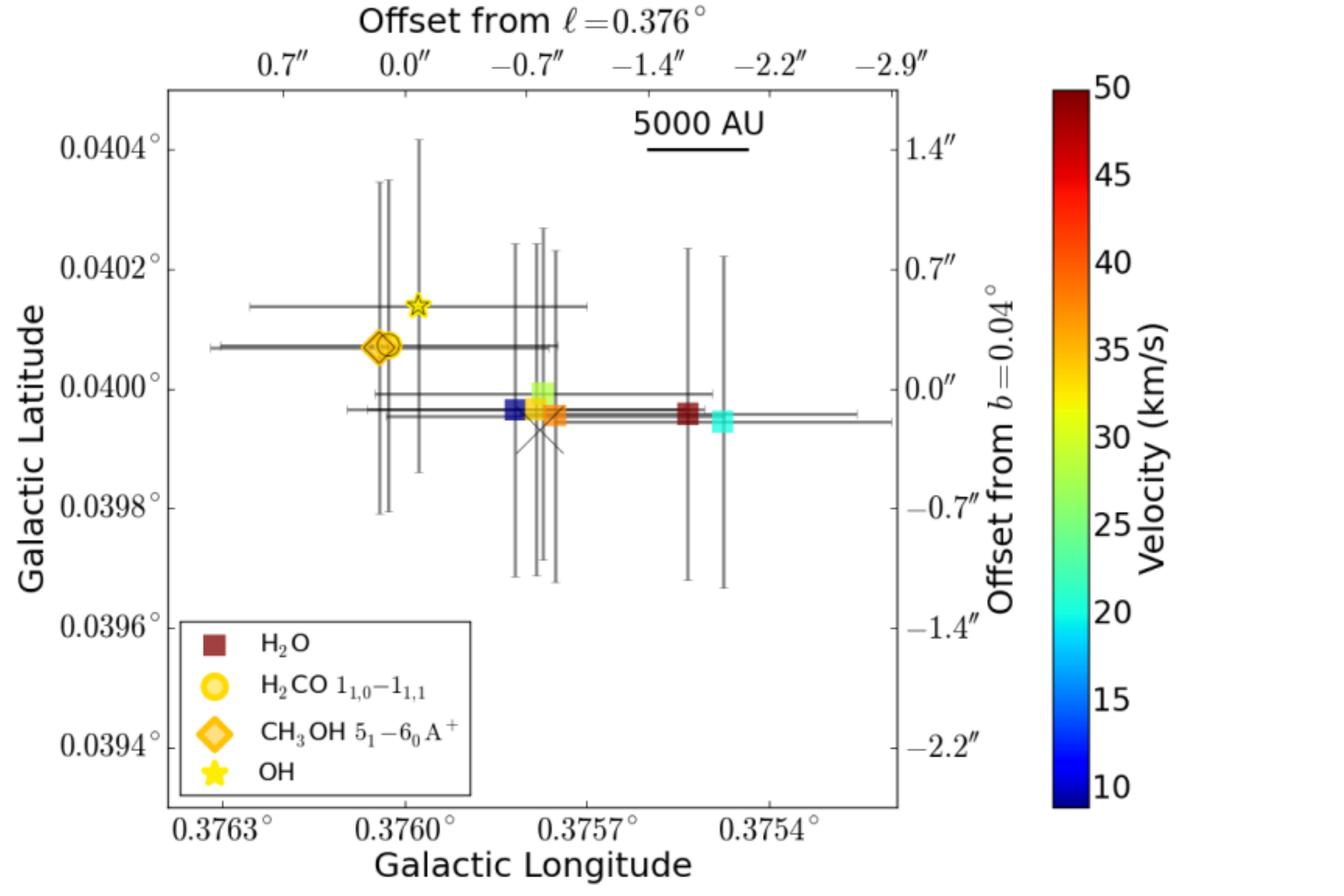}
{Overview of the detected masers colored by velocity.  The positional errors
on the SiO and \methanol \sevensix measurements are much larger than for the other data
sets because the measurements are low signal-to-noise from single-dish
observations, yet they are still likely to be underestimated (see Section
\ref{sec:othermasers}).  The gray boxes show the pixel size from the Mopra
observations of these lines.  (b) is a zoomed-in version of (a) focusing on the
interferometer observations.  The large X marks the centroid location of the
SMA-detected `core' (Walker et al in prep).
}
{fig:masers}{1}{3.5in}

Since the source is spatially and spectrally unresolved, this brightness
temperature is a lower limit.  If the true emitting area is 200 au, such as in
the \citet{Hoffman2007a} Sgr B2 maser spots, the true surface brightness is
$T_B=10^{7.4}$ K.  If the line is narrower than our upper limit of $\sigma<1.3$
\kms, it may be even brighter.

A literature search revealed that both a Class-II \methanol \fivesix
(6.67 GHz) maser and \water and OH masers have been detected toward Cloud C
\citep{Caswell1998a,Argon2000a,Pestalozzi2005a,Caswell2009a,Caswell2010a,Walsh2011a,Walsh2014a}.
We have measured the position of the 6.67 GHz \methanol maser from our own data, and it coincides
with the \formaldehyde maser in position to well within the statistical fit errors,
much less than the absolute positional uncertainty ($<0.1\arcsec$).
There is a water maser within 1 \kms of the \formaldehyde line, and the
brightest water maser is separated by only 4 \kms, so these may arise from the
same region; these \water masers are coincident with the \formaldehyde masers
to within the systematic pointing errors.  The OH and \methanol masers are also
within about 1 \kms of the \formaldehyde line.

We searched the \citet{Jones2013a} Mopra 7mm survey of the Class I \methanol
\sevensix (44.069476 GHz) line for emission and found a weak,
spatially unresolved line with peak brightness 0.06 K (0.5 Jy) at the position
and velocity of the \formaldehyde maser.  Assuming the emission comes from
$<1\arcsec$ on the sky, as is observed in the \formaldehyde line, the true
brightness must be $>300$ K, which suggests that this transition is masing.
However, \citet{Chambers2011a} observed this transition with the EVLA and
reported a nondetection with a sensitivity of 70 mJy/beam, so further
investigation of this line is warranted.

We also searched the \citet{Jones2013a} data for the SiO v=1 and v=2 J=1-0
lines (43.122079 and 42.820582 GHz).  We have clearly detected spatially
unresolved emission in both lines at $\sim64$ \kms at the position of Cloud C.
The detection of vibrationally excited SiO is a strong indication that these
are indeed masing transitions.  Figure \ref{fig:masers} shows that there is a
position offset between the Mopra-detected SiO and \methanol 44 GHz masers and
the ATCA-detected masers.  This is most likely because the centroid errors from
the fit to the Mopra moment-0 images are underestimated; there are systematic
errors in the Mopra maps (`streaking' artifacts) that affect sub-resolution
centroiding.

The SiO masers are offset by $\sim15-20$ \kms from most of the other lines, but
their velocities lie within the full range of the \water masers.  This
difference suggests that the \formaldehyde and \methanol and some of the \water
maser points trace a central protostellar core or disk, while the high-velocity
\water and SiO lines may trace part of an outflow or some other structure.

Finally, we searched the \citet{Jones2012a} Mopra 3mm survey for SiO v=1 J=2-1
86.243 GHz emission, but did not detect any, with a $3-\sigma$ upper limit of
96 mK or 0.89 Jy.  Given the detection of the 1-0 line at 0.56 Jy, this
nondetection is not surprising.

The spectral resolution of the Mopra data is 3.6 \kms, which is close to the 
FWHM of the measured lines.  Given the limited signal-to-noise ratio in these
data, the lines are consistent with being spectrally unresolved.

Table \ref{tab:measurements} shows the measured maser lines toward Cloud C,
including archival data.  Figure \ref{fig:masers} shows the maser spots in
position/velocity space.

\subsection{Comparison to other \formaldehyde and SiO maser sources}

To provide context, we summarize the other detected SiO and \formaldehyde
masers in the Galaxy.  The Sgr B2 maser region, the only other one to have
both lines detected as masers, shows a velocity offset between SiO
and \formaldehyde similar to the offset in Cloud C.

\textit{Orion KL}:
The Orion KL SiO masers are well-studied with a long VLBI monitoring program
showing that these lines trace the rotating base of an outflow driven by a disk
wind \citep{Goddi2009a,Greenhill2013a}.  The \water and SiO masers are closely
matched in velocity and generally spatially close: their emission centroid
is on the same position \citep{Greenhill2013a}.  No \formaldehyde maser
emission is seen toward Orion KL; the \formaldehyde \oneone emission seen there
is thermal with a peak $T_B\approx40$ K \citep{Mangum1993b}.

\textit{Sgr B2 (M)}:
There is only one SiO maser spot in Sgr B2, located near \citet{Mehringer1994b}
\formaldehyde Source C (not to be confused with Cloud C, the topic of this
paper).  The Sgr B2 \formaldehyde maser C is peculiar even among the Sgr B2 masers
in that the emission appears to be spatially and spectrally resolved, whereas
in other \formaldehyde masers in Sgr B2, the emission is unresolved.  While
this might normally hint at thermal emission processes, the high brightness
temperature ($T_B \sim 7300$ K) indicates instead that there must be multiple
unresolved maser spots within the source.
\citet{Zapata2009c} note that the SiO maser is shifted by about 20~\kms from
the cloud rest velocity, $v_{SiO}=87$~\kms,  while $v_{cloud}\sim60$~\kms; by
contrast, the \formaldehyde maser is near the cloud velocity or somewhat
blueshifted, with $v_{\formaldehyde} < 55$~\kms \citep{Mehringer1994b}.  The
remaining \citet{Mehringer1994b} \formaldehyde maser spots do not have
corresponding SiO masers.

\textit{W51 North}:
W51 North is a high-mass YSO that exhibits a rich spectrum of \ammonia
masers but has no centimeter continuum source \citep{Henkel2013a,Goddi2015a}.
It is detected in SiO at approximately the cloud rest velocity
\citep{Zapata2009c}, but is not detected in \formaldehyde~\oneone emission 
with an upper limit $<5$ mJy in a 1~\kms channel (Ginsburg et al in prep).

\textit{Other \formaldehyde sources}:
The remaining high-mass star-forming regions with \formaldehyde maser
detections in \citet{Araya2007b} and \citet{Araya2008a} do not have known
corresponding SiO masers
(G29.96-0.02, NGC 7538, G23.01-0.41, G25.38-0.18, G23.71-0.20, IRAS 18566+0408).  
However, of these, only NGC 7538 has been searched
for SiO masers \citep{Zapata2009a}. Out of the Araya and Zapata surveys, which
each searched $\sim60$ sources, there were only 12 sources common to both
samples.

\section{Discussion}
Out of the now eight known \formaldehyde maser-containing regions in the
Galaxy, two are in the CMZ. These two regions, Cloud C and Sgr B2, are
the only dense clouds in the CMZ with confirmed ongoing accretion onto a
high-mass YSO\footnote{Sgr C also shows some hints of accretion onto high-mass YSOs via
detected outflows
\citep{Kendrew2013a} and a 6.67 GHz \methanol maser \citep{Caswell2010a},
but it has not yet been searched for \formaldehyde masers.  The ultracompact
HII regions in Sgr B1 and the 20 and 50 \kms clouds appear to be more evolved
\citep{Mills2011a} and may no longer be accreting.  Cloud E contains a compact
molecular core and a 6.67 \methanol GHz maser \citep[Walker et al in
prep,][]{Caswell2010a}, but no \formaldehyde maser is detected.  There is one 
more 6.67 GHz \methanol maser source south of G0.253+0.016 that may be an isolated
site of high-mass star formation.}.
Cloud C and Sgr B2 (M) are also the only \formaldehyde maser sources with
corresponding SiO maser detections and vice versa, though the sample of regions
explored in both tracers is small.

This high detection rate of masers in star-forming regions within the CMZ,
despite limited statistical information, suggests that \formaldehyde
masers may be an efficient tracer of high-mass star formation in extreme
environments.
By contrast, extensive surveys have shown that the occurrence of \formaldehyde
masers in ``normal'' high-mass star-forming regions in the Galaxy is very low,
$<2\%$, or 1 of 58 sources in a large survey
\citep{Araya2004a,Araya2007b,Araya2008a,Ginsburg2011a,Ginsburg2015a}.

Given the overall rarity of both SiO masers and \formaldehyde masers toward
star-forming regions and their apparent prevalence in such regions within the
CMZ, is there something different about how high-mass star formation proceeds
in the CMZ?  Physical conditions on parsec scales are known to be very
different from those in the disk, with greater turbulent velocity dispersion
\citep{Shetty2012a}, higher gas temperatures \citep{Ao2013a,Ginsburg2015b},
higher dust temperatures (Battersby et al. in prep), higher pressure
\citep{Kruijssen2013a}, and widespread emission from shock tracers like
(thermal) SiO and HNCO \citep{Jones2012a}.  However, maser emission comes from
very small regions $\lesssim100$\,AU, so why should these parsec-scale
differences affect the forming stars?

One possibility is that these rare masers trace a very short period in the
lifetime of the forming high-mass YSO.  Both masers may
trace either an outflow or a disk \citep{Eisner2002a,Goddi2009b}, but the
conditions that allow them to mase may in either case last for a very short
time.  In this scenario, the presence of two such regions in the CMZ indicates
that there is currently an ongoing burst of star formation.  

Another possibility, which is more closely related to the driving mechanism of the
masers, is that high abundance of these species in the CMZ continues from
parsec scales down to $\sim100$ au scales.  While \formaldehyde is abundant
throughout the ISM and can be produced in the gas phase, its abundance is
greatly increased when grain surfaces are heated and ices sublimated.  SiO is
expected to rapidly deplete from gas into dust in the ISM, but its prevalence
throughout the CMZ indicates that there is a great deal of dust processing
releasing it into the gas phase.  \methanol is also prevalent throughout the
CMZ, and the high abundance required to produce detectable maser emission
implies it is formed on icy grain surfaces and subsequently sublimated, so its
presence is again an indication of grain destruction or heating.  The
widespread higher gas-phase abundances of these species may allow all high-mass
YSOs to go through a phase of \formaldehyde and SiO maser emission in the CMZ,
while in ``normal'' Galactic disk star formation, they cannot.

This abundance-based argument would also favor the formation of water masers.
However, \water masers are underabundant in the CMZ compared to the Galactic
disk, though they are present in both Sgr B2 and Cloud C \citep{Walsh2014a}.
\citet{Longmore2013b} note that the ratio of \water masers to thermal \ammonia
emission is orders of magnitude lower in the CMZ than the rest of the Galaxy.
By contrast, there is a (statistically weak) excess of \formaldehyde and SiO
masers.  If the \water masers come primarily from outflows, it may be that the
greater turbulence in the CMZ prevents an adequate path length from being
assembled in CMZ gas.  Another possibility is that existing \water maser
observations are not sensitive enough, and a population of lower-luminosity
maser sources has so far been missed \citep{Urquhart2011a}.  Furthermore, the
higher pressure and more turbulent CMZ environment means that prestellar cores
should form with higher densities \citep{Kruijssen2014c,Rathborne2014b}, which
may modify which masers are favored.

\section{Conclusion}

Cloud C in the CMZ dust ridge, a high-mass star-forming region, is revealed as
one of the most maser-rich sites in the Galaxy.  We have reported new
detections of \formaldehyde \oneone 4.82966 GHz, \methanol \sevensix
44.069476 GHz, SiO v=1 J=1-0 43.122079 GHz, and SiO v=2 J=1-0 42.820582 GHz
masers.  This cloud had not previously been identified as a maser-rich region
because both  the region and its accompanying masers are faint at all
wavelengths compared to neighboring Sgr A and Sgr B2. However, as a maser-rich
region, it should prove a useful ground for testing maser mechanisms in unusual
masing transitions.

The detection of these masers raises questions about star formation in the CMZ.
It is likely that CMZ chemistry and turbulence are different enough from the
Galactic disk that masers in the CMZ trace different stages of star formation.
Further surveys for rare maser lines toward star forming regions in the inner
few hundred parsecs are needed to confirm this speculation.  Additionally,
further searches for both SiO and \formaldehyde masers toward a consistent set
of target regions would help determine how unique the association between these
masers really is.

\begin{acknowledgements}
This work made use of MIRIAD \citep{Sault1995a}, astropy
\citep{Astropy-Collaboration2013a}, aplpy (\url{aplpy.readthedocs.org}),
pyspeckit \citep{Ginsburg2011c}, ds9 (\url{ds9.si.edu}), astroquery
(\url{astroquery.readthedocs.org}), splatalogue (\url{splatalogue.net}) and its
member catalogs \citep{Pickett1998a,Muller2005b}, and the radio-astro-tools
toolkit (\url{radio-astro-tools.github.io}).  JMDK is funded by a Gliese
Fellowship.  This letter is based on data from ATCA project C3045.
\end{acknowledgements}

\bibliographystyle{aa}

\begin{appendix}
\section{Spectra \& Summary Table}
We show the extracted spectra in Figure \ref{fig:spectra} and the summary of
the detected maser lines in Table \ref{tab:measurements}.

\Figure
{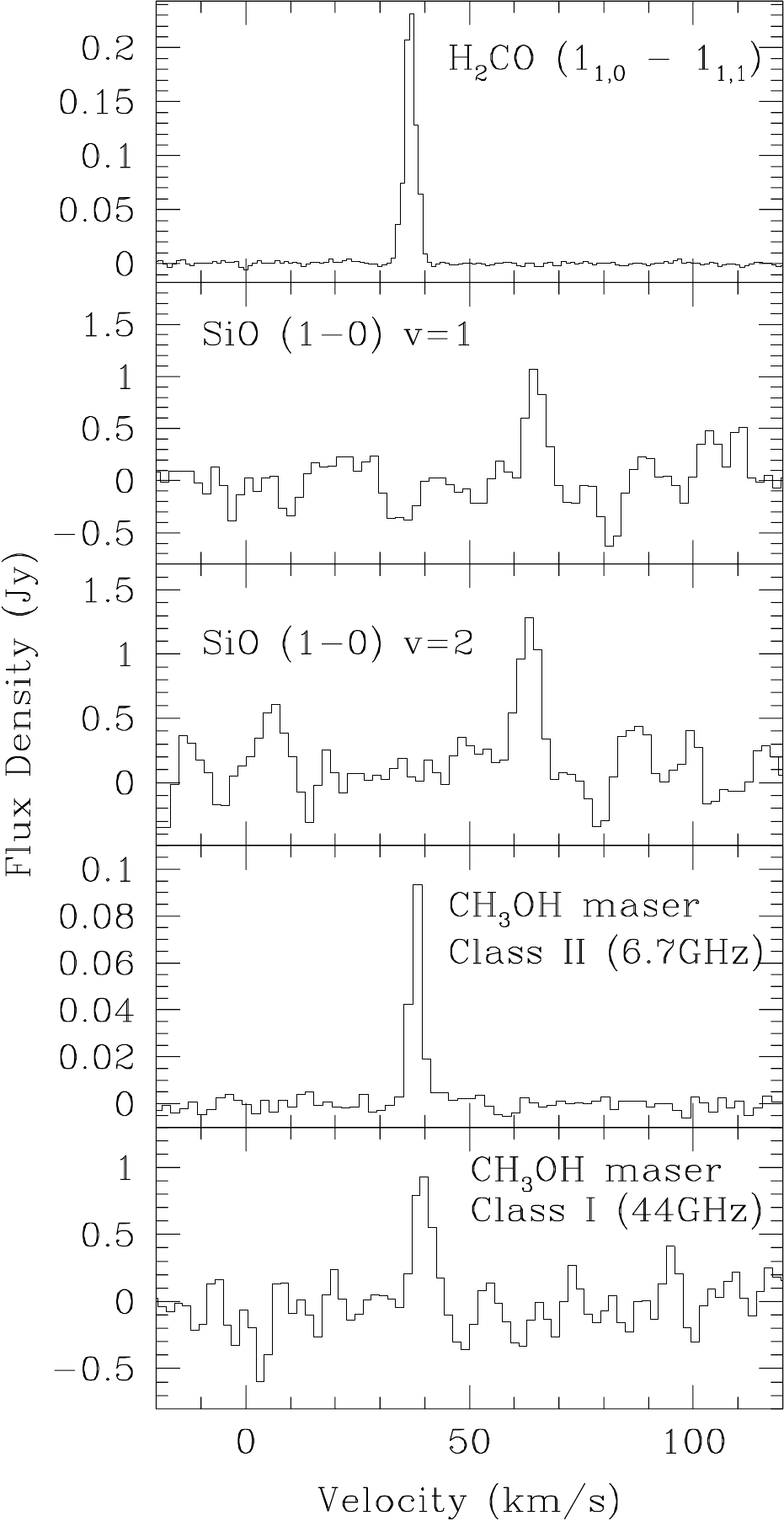}
{Spectra of each of the measured lines.  The Mopra data, both SiO lines and the
\methanol \sevensix 44 GHz Class I line, have spectral resolution 3.6 \kms and
channel spacing 1.8 \kms.  The ATCA data have spectral resolution 1.9 and 1.4
\kms and beam shapes 4.85\arcsec $\times$ 1.49\arcsec and 9.4\arcsec $\times$
0.46\arcsec for the \formaldehyde and \methanol lines, respectively.}
{fig:spectra}{1}{3.0in}

\begin{table*}[htp]
\caption{Maser line parameters}
\begin{tabular}{cccccccc}
\label{tab:measurements}
Line & $\ell$ & $b$ & $\sigma(\ell)$ & $\sigma(b)$ & $v_{LSR}$ & $\sigma(v_{LSR})$ & Measurement \\
 & $\mathrm{{}^{\circ}}$ & $\mathrm{{}^{\circ}}$ & \arcsec & \arcsec & $\mathrm{km\,s^{-1}}$ & $\mathrm{km\,s^{-1}}$ &  \\
\hline
CH$_3$OH \sevensix & 0.38270 & 0.03630 & 6.8 & 5.4 & 39.6 & 0.4 & This Work \\
H$_2$CO $1_{1,0}-1_{1,1}$ & 0.37603 & 0.04007 & 1 & 1 & 36.7 & 0.01 & This Work \\
SiO J=1-0 v=1 & 0.37520 & 0.03280 & 5.04 & 4.3 & 64.8 & 0.4 & This Work \\
SiO J=1-0 v=2 & 0.37380 & 0.03940 & 4.3 & 6.8 & 63.2 & 0.4 & This Work \\
CH$_3$OH \fivesix & 0.37604 & 0.04007 & 1 & 1 & 38.0 & 0.05 & This Work \\
H$_2$O G000.375+0.042\_A & 0.37545 & 0.04153 & 1 & 1 & 78.8 & - & Walsh 2014 \\
H$_2$O G000.376+0.040\_A & 0.37582 & 0.03996 & 1 & 1 & 9.5 & - & Walsh 2014 \\
H$_2$O G000.376+0.040\_B & 0.37548 & 0.03995 & 1 & 1 & 24.4 & - & Walsh 2014 \\
H$_2$O G000.376+0.040\_C & 0.37577 & 0.03999 & 1 & 1 & 32.3 & - & Walsh 2014 \\
H$_2$O G000.376+0.040\_D & 0.37578 & 0.03997 & 1 & 1 & 37.3 & - & Walsh 2014 \\
H$_2$O G000.376+0.040\_E & 0.37575 & 0.03995 & 1 & 1 & 40.4 & - & Walsh 2014 \\
H$_2$O G000.376+0.040\_F & 0.37553 & 0.03996 & 1 & 1 & 52.5 & - & Walsh 2014 \\
OH & 0.37598 & 0.04014 & 1 & 1 & 36.0 & - & Caswell 1998 \\
\hline
\end{tabular}
\par
Statistical errors on the fit position are given for the single-dish data, and an assumed lower-limit systematic error of 1\arcsec is given for each of the interferometric observations.
\end{table*}

\end{appendix}

\end{document}